%% file: main.tex
\documentclass[acmtog,nonacm]{acmart}

\usepackage{booktabs} 

\input{sec/_preamble.tex}

\input{sec/_math.tex}
\inlinefigstrue

\usepackage[ruled]{algorithm2e} 
\usepackage{subcaption}

\newcommand{\name}{CurveCodec~2\xspace}
\newcommand{\prevname}{CurveCodec\xspace}

\acmJournal{TOG}
\acmYear{2026}
\copyrightyear{2026}

\begin{document}

\title{CurveCodec 2: Skeleton-Agnostic Animation Compression with a Learned Entropy Model}

\author{Mingyi Shi}
\orcid{0000-0002-5180-600X}
\affiliation{
  \institution{The University of Hong Kong}
  \city{Hong Kong}
  \country{Hong Kong}
}
\email{irubbly@gmail.com}

\author{Huancheng Lin}
\orcid{0000-0003-4446-1442}
\affiliation{
  \institution{The University of Hong Kong}
  \city{Hong Kong}
  \country{Hong Kong}
}
\email{lamws233@gmail.com}

\author{Taku Komura}
\authornote{Corresponding author.}
\orcid{0000-0002-2729-5860}
\affiliation{
  \institution{The University of Hong Kong}
  \city{Hong Kong}
  \country{Hong Kong}
}
\email{taku@cs.hku.hk}

\input{sec/0_abs}


\begin{teaserfigure}
    \centering
    \includegraphics[width=\textwidth]{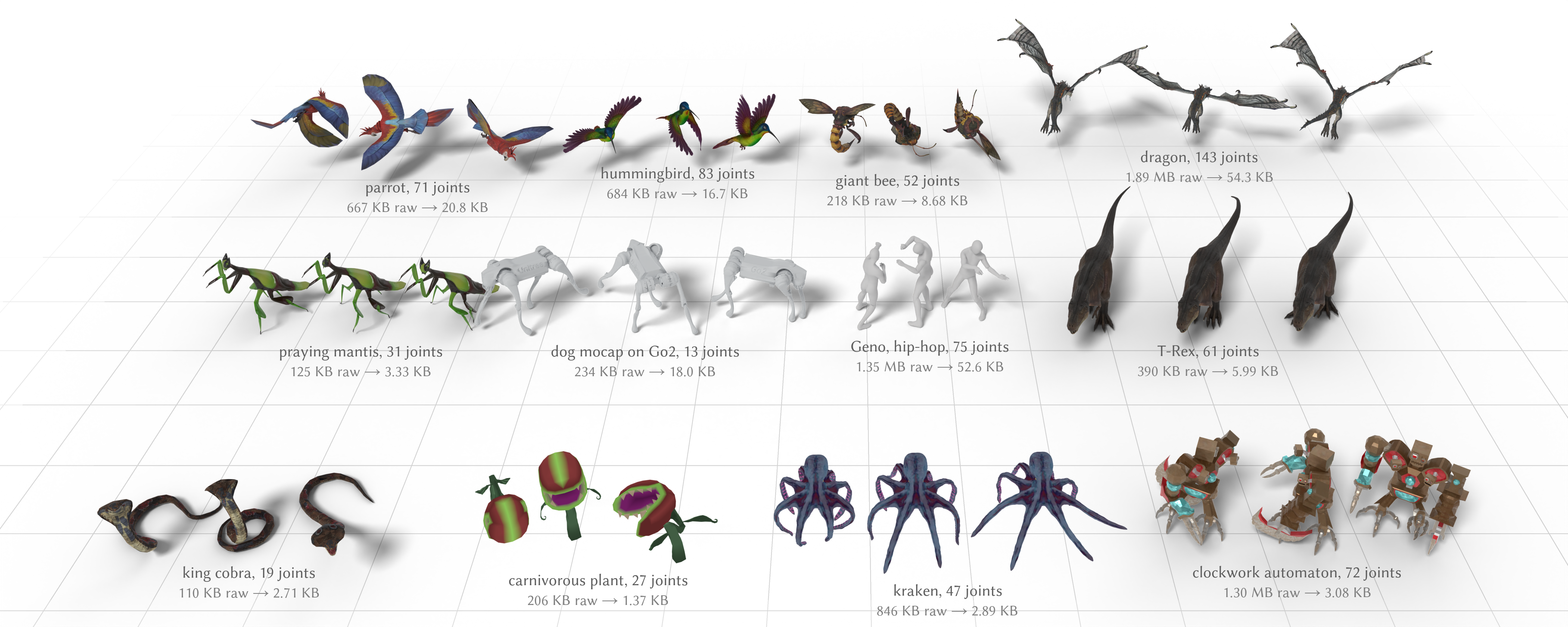}
    \caption{\capTeaser}
    \Description{\descTeaser}
    \label{fig:teaser}
\end{teaserfigure}

\maketitle

\input{sec/1_intro}
\input{sec/2_related}
\input{sec/3_prelim}
\input{sec/4_method}
\input{sec/5_analysis}
\input{sec/6_exp}
\input{sec/7_conclusion}

\bibliographystyle{ACM-Reference-Format}
\bibliography{main.bib}

\end{document}

%% file: sec/_preamble.tex
\usepackage{graphicx}
\usepackage{amsfonts}
\usepackage{amsmath}
\usepackage{pifont}
\usepackage{booktabs}
\usepackage{xcolor}
\usepackage{multirow}
\usepackage[normalem]{ulem}
\usepackage[super]{nth}
\usepackage{cleveref}
\usepackage{colortbl}
\usepackage{tabularray}
\usepackage{listings}
\usepackage{tikz}
\usetikzlibrary{positioning,arrows.meta,fit,calc,backgrounds}

\newif\ifdraft
 \drafttrue
\ifdraft
\newcommand{\myc}[1]{{\color{blue}\textbf{Mingyi:} #1}}
\newcommand{\xlc}[1]{{\color{magenta}\textbf{Xuelin:} #1}}
\newcommand{\tkc}[1]{{\color{cyan}\textbf{Taku:} #1}}
\newcommand{\ruic}[1]{{\color{green}\textbf{Rui:} #1}}

\else
\newcommand{\myc}[1]{}
\newcommand{\xlc}[1]{}
\newcommand{\tkc}[1]{}
\newcommand{\ruic}[1]{}

\fi

\usepackage{xspace}

\newcommand{\keep}[1]{}
\newcommand{\old}[1]{}

\usepackage{url}
\usepackage{hyperref}

\newif\ifinlinefigs
\inlinefigsfalse
\input{sec/_figcaptions.tex}

%% file: sec/_figcaptions.tex
\newcommand{\capTeaser}{\name{} is a skeleton-agnostic codec for skeletal animation. Twelve rigs from the example set of our online demo, each at three frames of its clip as decoded by \name{} at ACL's default precision of $0.01$\,cm, with the clip's raw size and the size of our stream under each. One model, trained once, codes all of them without retraining. On the held-out test side \name{} needs $0.37\times$ ACL's bytes at this precision, from a bit-exact bitstream that decodes on one CPU core. Project page, online demo, and supplementary material: \url{https://rubbly.cn/publications/curvecodec/}}
\newcommand{\descTeaser}{Twelve rigged characters of different skeletons, birds, a bee, a dragon, a mantis, a robot dog, a human, a dinosaur, a snake, a plant, a kraken and a machine, each in three poses of its decoded clip, in three rows on a grid floor drawn in perspective.}

\newcommand{\capPipeline}{Overview of \name{}. (a) The lossy stage runs once per clip in the encoder. Every animated sub-track becomes a curve in the log map; a closed loop then proposes two kinds of moves, growing a track's quantization step or dropping one of its keys, decodes the candidate, evaluates the shell error of the affected subtree through forward kinematics, and accepts the move only if the error contract still holds. The skeleton's geometry is used here and nowhere else; the lossless stage reads only each joint's depth in the hierarchy. (b) The lossless stage codes the resulting integers. A fixed predictor turns them into residuals, and a small learned model, the only learned component (a causal transformer in the product, an MLP under the max gate), reads features of each curve's own past and outputs a mixture distribution that drives the rANS coder. The decoder replays the same model in lockstep over all curves, so it reproduces every probability exactly, and reconstructs the clip by Catmull--Rom interpolation between keys.}
\newcommand{\descPipeline}{Two-panel block diagram: a closed loop of step growth and key removal verified by forward kinematics, and an entropy coder with a causal transformer shared by encoder and decoder.}

\newcommand{\capTwoStage}{The two-stage view of animation compression, with the estimated size of our $906$-hour corpus after each stage at $p = 0.01$\,cm (extrapolated from the test side). (a) Deterministic preprocessing shared by ACL and \name{}: dropping the absent scale sub-tracks and folding sub-tracks that do not move. (b) ACL quantizes the remaining sub-tracks with per-segment range reduction and fixed bit widths ($32.6$\,GB); \name{} codes them as curves with a closed-loop quantizer, RD-selected keys verified through forward kinematics, prediction, and a learned entropy coder ($11.9$\,GB). The $870$\,KB model is fixed after training and shared by all clips; it is not counted in the stream.}
\newcommand{\descTwoStage}{Block diagram of shared preprocessing followed by two rows, ACL's stages and ours, with the corpus size after each.}

\newcommand{\capAclBudget}{Where ACL's bits go on the curated set. (a) Bits per joint-sample after each ACL stage: almost all of the $17.7\times$ ratio comes from folding sub-tracks that do not move ($2.84\times$) and from the two-level range reduction with variable bit widths ($3.74\times$). (b) Entropy of ACL's own quantized symbols relative to the bits ACL spends, for zeroth- to second-order models: an entropy coder with these models on top of ACL saves at most $16$ to $20\,\%$, so a substantially smaller stream needs a different quantizer, not just a better back end.}
\newcommand{\descAclBudget}{Horizontal bars of bits per joint-sample through ACL's stages, and vertical bars of the entropy of ACL's symbols.}

\newcommand{\capRD}{Rate against precision on the held-out test side ($4{,}472$ clips, $20.1$\,h, $33$ datasets). Left: bits per joint-sample. Right: our bytes divided by ACL's on the same clips. Under the max gate (orange), which holds every joint within $5\,\%$ of the larger of $p$ and ACL's maximum, \name{} needs $0.41\times$ to $0.22\times$ ACL's bytes for $p \leq 0.1$\,cm; under the iso-mean criterion (blue) the ratio keeps falling to $0.07\times$ at $1$\,cm. All ratios include the fallback for clips that fail their contract. At $0.3$ and $1$\,cm ACL misses its own precision on part of the data (dashed lines, whole test side); hollow markers on dotted lines repeat the ratio on the clips where ACL keeps at least $99\,\%$ of its joint-samples within $p$.}
\newcommand{\descRD}{Two line plots: bits per joint-sample and the ratio to ACL as a function of the precision, for ACL and both of our error contracts.}

\newcommand{\capKeys}{How the encoder spends distortion. (a) Fraction of samples kept as keys by joint depth (five development clips, an earlier version of the encoder): the root, whose error reaches every descendant, keeps the most keys, and leaves the fewest. (b) The keys the encoder keeps fall from $79\,\%$ to $16\,\%$ of the samples as $p$ loosens, yet at $p \leq 0.1$\,cm more than three quarters of the remaining gaps hide at most two samples. (c) The marginal distortion an encoder move adds per bit it saves, at $p = 0.1$\,cm: removing a key is $5$ to $10\times$ cheaper than growing the quantization step or adding a dead zone, which is why the encoder prefers to omit samples once the precision leaves room to.}
\newcommand{\descKeys}{Three panels: grouped bars of keys kept by depth, line plots of keys kept and gap-length shares against precision, and bars of distortion per bit for three encoder moves.}

\newcommand{\capOracle}{Interpolation from the decoded neighbourhood gains little on the gaps an RD encoder leaves. A nonparametric nearest-neighbour oracle searches every window of the training corpus for the best match of a gap's own-curve context ($k$ is cubic error divided by oracle error; $k > 1$ is better than cubic). On random gaps (grey) it beats cubic interpolation by up to $1.37\times$ and keeps improving with the corpus. On the gaps the RD encoder actually leaves (orange) it is no better than linear interpolation and is flat in the corpus size over the range we could measure.}
\newcommand{\descOracle}{Line plot of the oracle ratio against corpus size for random gaps, rising above one, and for RD-selected gaps, flat below one.}

\newcommand{\capPerDataset}{Our bytes divided by ACL's on each dataset of the held-out test side, iso-mean criterion, at three precisions, sorted by the ratio at $0.01$\,cm. No dataset is larger than ACL. The gain is largest on long, high-rate captures (large online mocap, HiPHI, BEAT) and smallest on short, low-rate clips of small skeletons (HumanAct12, Edinburgh, BFA). The dog (orange) is a species absent from training.}
\newcommand{\descPerDataset}{Grouped bar chart of the compression ratio relative to ACL for 33 datasets at three precisions.}

\newcommand{\capStorage}{Estimated storage of the full 906-hour corpus ($16.17$\,G joint-samples), extrapolated per dataset from the test side; the full corpus was not encoded. Our bars use the mean-gate product codec after the fallback. Log scale.}
\newcommand{\descStorage}{Bar chart, log scale, of the corpus size in BVH text, raw floats, ACL and our format at three precisions.}

\newcommand{\capQualitative}{Decoded motion on the held-out test side. Each half-row shows the ground-truth mesh, the skeletons decoded by ACL, by \name{} under the max gate, and by \name{} under the mean gate, each overlaid on the ground-truth skeleton (blue; the decoded skeleton is orange and thinner, so where they coincide the overlay reads pink), and the mesh driven by our decoded motion. Labels give the clip's compression ratio against ACL's raw size and its maximum shell error. Top two rows: $p = 0.01$\,cm, where all three codecs are visually exact and \name{} needs $2$ to $3\times$ fewer bytes than ACL. The dog capture (held out of training) has no mesh of its own; its mesh panels show a Go2 robot for orientation only and its skeleton panels are placed approximately. Bottom row: $p = 1$\,cm on the same humanoid clip, with insets on the joint that moves the most; ACL reaches $39\times$ and \name{} $172\times$ (max gate, ACL's worst case within tolerance) or $182\times$ (mean gate, same mean error), and the largest joint displacement in the shown frame is $0.46$ to $0.63$\,cm for all three.}
\newcommand{\descQualitative}{Rows of characters showing ground truth, ACL and our decoded skeletons overlaid.}

\newcommand{\capFailure}{The weakest case: a $192$-sample, $20$\,Hz clip of a $27$-joint skeleton (HumanAct12) at $p = 0.01$\,cm. All codecs are visually exact, but \name{} saves only a quarter of ACL's $23.8$\,KB ($0.74\times$ to $0.75\times$), against $0.37\times$ on the whole test side: the per-clip header and the cold start of the entropy model are amortized over few samples, and at $20$\,Hz few samples can be omitted.}
\newcommand{\descFailure}{A short clip with its decoded skeleton and per-joint bit costs.}

%% file: sec/_math.tex
\newcommand{\realnum}{ \mathbb{R} }
\newcommand{\integers}{ \mathbb{Z} }

\newcommand{\nbjoint}{ J }            
\newcommand{\nbframe}{ N }            
\newcommand{\motion}{ M }             
\newcommand{\rotation}{ \textbf{q} }  
\newcommand{\translation}{ \textbf{t} }
\newcommand{\logrot}{ \textbf{v} }    
\newcommand{\parent}{ \pi }           

\newcommand{\shell}{ \delta }         
\newcommand{\err}{ e }                
\newcommand{\budget}{ m_p }           
\newcommand{\aclmax}{ \bar{e}^{\mathrm{ACL}} }
\newcommand{\aclmean}{ \mu^{\mathrm{ACL}} }
\newcommand{\gcap}{ \rho }            
\newcommand{\gtol}{ \tau }            
\newcommand{\gguard}{ \gamma }        
\newcommand{\gfloor}{ \phi }          

\newcommand{\step}{ s }               
\newcommand{\gridres}{ G }            
\newcommand{\gidx}{ i }               
\newcommand{\qint}{ n }               
\newcommand{\rdom}{ R }               
\newcommand{\chain}{ c }              

\newcommand{\keys}{ \mathcal{K} }
\newcommand{\dist}{ \Delta D }
\newcommand{\rate}{ \Delta R }
\newcommand{\lagr}{ \lambda }
\newcommand{\gbits}{ \beta }

\newcommand{\resid}{ r }
\newcommand{\pred}{ \hat{\qint} }
\newcommand{\feat}{ \textbf{f} }
\newcommand{\ctxwin}{ W }
\newcommand{\nbmix}{ K }
\newcommand{\entnet}{ \mathcal{P}_\theta }

\DeclareMathOperator{\rint}{rint}
\DeclareMathOperator{\Log}{Log}
\DeclareMathOperator{\Exp}{Exp}
\newcommand{\norm}[1]{\lVert #1 \rVert}

%% file: sec/0_abs.tex
\begin{abstract}

Skeletal motion is stored as every joint's transform at every frame, yet most of it is implied by the body rather than by what the motion is about. Compression is one way to ask what a motion must still say once the body is known, and a production codec is where that question has to be answered for any skeleton, with a stated bound on the error. Our earlier codec, \prevname{}, reconstructed each joint curve from sparse anchors with a learned prior. It matched the mean error of ACL, the production library of modern game engines, but not its worst case, and it counted its payload as floating-point values rather than as bits.
Here we ask where the redundancy of skeletal motion lies, and which part of a codec a learned model should take over. A measured anatomy of ACL and a series of controlled experiments give three answers. At production precision the largest saving comes from predicting each quantized curve from its own past, and the second from choosing per joint, in closed loop through the hierarchy, which samples \emph{not} to code. On the gaps such an encoder leaves, interpolation has little left to find: a nearest-neighbour oracle with millions of training samples as memory is no better than linear interpolation there, and none of the learned in-betweeners we tried paid for itself. What a network does learn is the distribution of the residuals the codec must send.
\name{} puts these findings into one codec. Every sub-track becomes a curve, rotations in the log map and translations in centimetres, quantized in closed loop and thinned to rate--distortion-selected keys, and its prediction residuals are entropy-coded under a small learned model whose integer inference is bit-exact across platforms. Two contracts are verified on every decoded clip: ACL's own worst case per joint, within a stated tolerance, or ACL's mean error per clip.
On a held-out test side of $4{,}472$ clips from $33$ datasets, \name{} needs $0.37\times$ ACL's bytes at ACL's default precision of $0.01$\,cm under the worst-case contract and $0.22\times$ at $0.1$\,cm under the mean contract. It decodes whole clips on one CPU core, typically in a fraction of a second, and transfers without retraining to a species absent from training.

\end{abstract}

%% file: sec/1_intro.tex
\section{Introduction}
\label{sec:intro}

A skeletal animation clip stores rotations and translations for tens to hundreds of joints at every sample, yet much of this is not what the motion is about. A performer decides what to do; the body, its morphology and its dynamics, determines most of how each joint gets from one place to the next. Compressing these data matters in production, where animation libraries face strict memory and streaming budgets, and in motion research, where corpora reach hundreds of hours and a model should spend its capacity on the decisions rather than on kinematics the body already implies. A codec is one way to probe this separation: what it must still send under a stated error bound is at least what the body did not predict. This is our motivation, not a claim the bit counts prove. If the body is only a prior, the representation should not be tied to one body: a production codec must serve arbitrary skeletons and motion without retraining, and it must bound its error, because a single visible pop on one joint is a defect regardless of how small the average error is.

ACL~\cite{frechette2023acl} is the industry standard. It folds the sub-tracks that do not move, reduces the range of the rest per clip and per segment, and quantizes each joint with one of a fixed set of bit widths under a per-joint error threshold. It decodes a pose in microseconds, but it models neither time nor the statistics of its own symbols. Learned motion encoders~\cite{ling2020character,yao2024moconvq,guo2024momask} take the opposite trade: their latent priors couple the representation to a training skeleton and domain, and they do not reach production precision.

Our previous work, \prevname{}~\cite{shi2026curvecodec}, placed the learned prior on the smallest skeleton-independent unit of motion, the local joint curve, and reconstructed each curve from sparse anchors with a shared transformer. It reached three times ACL's ratio at matched \emph{mean} error, but its worst case exceeded ACL's, its payload was counted in floating-point values rather than entropy-coded bits, and its decoder needed a GPU rather than a CPU core.

This paper asks a different question: \emph{where is the redundancy in skeletal motion, and which part of a codec should a learned model take over?} We answer it by measurement, treating the codec as an instrument: under an explicit error contract and an exact bit count, every candidate structure of motion either pays or does not. We dissect ACL stage by stage and test every place where a learned component could enter. Three findings shape the result.
First, ACL's ratio comes almost entirely from folding and from range reduction with variable bit widths. Its widths are nearly tight for independent samples, so an entropy coder on top of ACL's own symbols would gain little; a substantially smaller stream needs its own quantizer, one whose integers remain predictable in time.
Second, at production precision the largest saving comes from predicting each curve from its own past, and the next from choosing, per joint and in closed loop through the hierarchy, which samples to omit. Leaving a sample out is by far the cheapest way to spend error, and its share of the stream grows as the precision loosens, until the two are comparable at the loosest precision we use.
Third, on the gaps such an encoder leaves, interpolation has little left to find. A nearest-neighbour oracle with millions of training samples as memory, which clearly beats cubic interpolation on random gaps, is no better than \emph{linear} interpolation on the gaps the encoder actually leaves, and the learned in-betweeners we placed inside the codec did not pay. On our data, the in-betweener at the core of \prevname{} has little left to learn; what a network does learn is the \emph{distribution} of the residuals that the codec still has to send.

These findings lead to \name{}. Every animated sub-track is coded as a curve: rotations in the log map, which removes the precision floor of ACL's dropped quaternion component; a quantization step and a key set chosen in closed loop against the object-space error of the whole subtree; and prediction residuals coded with rANS under a small learned model of the curve's own past, a causal transformer or, in the worst-case configuration, an MLP. The model sees no joint identity, so the codec is skeleton-agnostic, and its integer inference is bit-exact on ARM and x86. Two contracts are verified on every decoded clip: ACL's own worst case per joint, within a stated tolerance, or ACL's mean error per clip. A clip that fails is re-encoded with a margin or stored as ACL's stream, and every number we report includes this fallback.

On a corpus of $906$ hours from $33$ datasets with a held-out test side, \name{} needs $0.37\times$ ACL's bytes at ACL's default precision of $0.01$\,cm under the worst-case contract, and $0.22\times$ at $0.1$\,cm under either contract; against ACL with its default keyframe stripping the ratios are slightly larger and the picture is the same. It decodes on a CPU, typically a clip in a fraction of a second on one core, but only whole clips: it is a storage and distribution format decoded at load time, not a replacement for ACL's constant-time sampling.

Our contributions are:
\begin{itemize}
    \item \textbf{A quantitative anatomy of motion redundancy}, measuring where a production codec's bits go and which structure a codec exploited in our hands: temporal predictability and, increasingly as the precision loosens, per-joint sample sparsity did; interpolation from the decoded neighbourhood, cross-joint structure, and transforms under a worst-case criterion did not.
    \item \textbf{\name{}, a closed-loop, skeleton-agnostic rate--distortion codec} under two explicit error contracts, with a stated fallback for the clips that fail.
    \item \textbf{A learned entropy model in the role where a network pays}: small, causal, and bit-exact across platforms, it models the distribution of the residuals the codec must send, the one place where, in our measurements, a network consistently earns its cost.
    \item \textbf{A corpus-scale benchmark against ACL}, with audits of ACL's configuration and of our own contracts, and same-core speed.
\end{itemize}

%% file: sec/2_related.tex
\section{Related Work}
\label{sec:related}

Skeletal animation compression is a long-standing problem in computer graphics. Early work focused on making animation data smaller and faster to decode in production systems; learned representations later showed that motion can be organized into compact spaces that support reconstruction and synthesis. Our work sits between the two: an explicit, verifiable codec in the production tradition, with a learned component placed where our measurements show that learning pays.

\subsection{Hand-engineered Animation Compression}

Before deep learning, animation compression removed temporal redundancy, inter-joint redundancy, and perceptually insignificant detail while keeping the motion faithful.
Mesh compression began with static connectivity and geometry~\cite{rossignac1999edgebreaker,karni2000spectral}; animated meshes were then compressed through their vertex trajectories, by principal components~\cite{alexa2000representing} or space-time prediction~\cite{ibarria2003dynapack}.
\citet{arikan2006compression} moved the problem to skeletal motion: joint angles were converted into virtual marker trajectories, fitted with cubic B\'ezier curves, and projected into clustered PCA subspaces.
Later work exploited low-dimensional structure with segment-level PCA and spline interpolation~\cite{liu2006segment}, quadratic B\'ezier break-and-fit curves~\cite{khan2016efficient}, wavelet coefficient selection~\cite{beaudoin2007adapting,lee2008efficient}, and wavelets guided by perception~\cite{firouzmanesh2011perceptually}; others indexed repeated motion patterns across a database~\cite{gu2009compression,lin2010adaptive}.
ACL~\cite{frechette2023acl} is the production standard. It prioritizes per-track processing and constant-time random access over compression ratio, and Sec.~\ref{sec:analysis:acl} quantifies where its bits go.
Our keys relate to keyframe extraction~\cite{lim2001key,halit2011multiscale} and curve simplification~\cite{barkowsky2000schematizing}, which choose samples from a local error or saliency criterion. \name{} instead chooses them by rate--distortion optimization against the object-space error of the whole subtree, and Sec.~\ref{sec:analysis:keys} shows that this closed loop, not the key representation itself, carries the gain. Unlike transform codecs, which we find lose to sparse keys under a worst-case criterion, \name{} keeps the time domain and predicts each sample from its own past.

\subsection{Learned Motion Representations and Codecs}

Motion manifolds and deep motion representations show that valid character motion occupies a much smaller space than raw joint rotations~\cite{holden2015learning,holden2016deep}. VAE-style priors support character control, action-conditioned synthesis, and pose estimation~\cite{ling2020character,petrovich2021action,rempe2021humor}; tokenizers with vector or residual quantization compress motion into discrete codes~\cite{yao2024moconvq,guo2024momask}; phase-based models organize motion by periodic timing~\cite{Starke2022DeepPhase,shi2023phasemp}; and Learned Motion Matching~\cite{holden2020learned} amortizes an animation database into networks. Cross-skeleton methods learn representations for retargeting~\cite{villegas2018neural,aberman2020skeleton,Lee2023SAMESM}.
These representations target synthesis, control, or transfer at pose level, not a decodable bitstream at a stated precision; SAME~\cite{Lee2023SAMESM} in particular learns an embedding shared across skeletons, which is a different goal from ours rather than a competing codec. The learned encoders measured in \prevname{}~\cite{shi2026curvecodec} under the shell metric reconstruct with errors two orders of magnitude above the $0.01$\,cm of a production codec, and the residual-VQ tokenizer we built has a precision floor of about $1.3\,\%$ of a track's RMS (Sec.~\ref{sec:analysis:net}); we do not extend these bounds to methods we did not measure.
Implicit neural representations fit signals with coordinate networks~\cite{park2019deepsdf,tancik2020fourier,sitzmann2019siren,dupont2021coin,chen2021nerv}, and NeMF~\cite{he2022nemf} applies the idea to kinematic motion. We find the same fidelity floor for per-window neural fields.
\prevname{}~\cite{shi2026curvecodec} is our direct predecessor. It made the learned prior skeleton-agnostic by working on single sub-tracks, and reconstructed each from a first-frame state, a Fourier descriptor, and extrema anchors with a masked transformer, in the spirit of learned in-betweening~\cite{harvey2020robust}. \name{} keeps the per-sub-track, skeleton-agnostic unit but moves the learned prior from the interpolator to the probability model, for reasons that Sec.~\ref{sec:analysis:interp} measures on our data.

\subsection{Entropy Coding and Learned Probability Models}

Lossless coding separates modelling from coding: an arithmetic coder~\cite{witten1987arithmetic} or an asymmetric numeral system~\cite{duda2013ans} spends $-\log_2 P$ bits per symbol, so the compression depends only on how well the model $P$ predicts the data. Predictive lossless coders such as LOCO-I / JPEG-LS~\cite{weinberger2000loco} and CALIC~\cite{wu1997calic} combine a fixed predictor with a context model of the residual, the structure \name{} adopts per curve.
Learned image compression~\cite{balle2017endtoend,balle2018variational,minnen2018joint} showed that much of the gain of a learned codec comes from the learned probability model, and that decoding with a network requires deterministic, integer inference to keep encoder and decoder in sync across platforms~\cite{balle2019integer}. General-purpose neural compressors use recurrent or transformer models as the probability model of a byte stream~\cite{goyal2019deepzip,bellard2021nncp,deletang2024language}.
Learned probability models have entered animation compression before: Chen et al.~\cite{chen2022latent} compress motion-capture sequences with a learned continuous-time latent model, and Qin et al.~\cite{qin2025multidescriptor} code mesh animations with a deep entropy model over vertex trajectories; neither holds an object-space worst-case bound through a skeleton. \name{} combines per-curve learned residual probabilities with hierarchy-aware quantization and key selection under explicitly verified object-space error constraints, and our experiments indicate that for skeletal motion the probability model is the role where a network pays, while the learned interpolators, transforms, tokenizers, and neural fields we tried did not pay for themselves.

%% file: sec/3_prelim.tex
\section{Preliminaries}
\label{sec:prelim}

\subsection{Skeletal Animation and Error}
\label{sec:prelim:motion}

A skeleton has $\nbjoint$ joints connected as a rooted tree; joint $j$ has parent $\parent(j)$. A clip of $\nbframe$ uniformly spaced samples stores, per joint and sample, a local rotation $\rotation_j(t)$ (a unit quaternion) and a local translation $\translation_j(t) \in \realnum^3$; BVH data carry no scale. Following ACL, the time series of one transform component of one joint is a \emph{sub-track}. Skeleton-agnostic means that the same codec and weights apply to any joint count and tree topology without retraining. For comparability with ACL we use its raw size of ten $32$-bit floats per joint-sample, but report rate mainly in \emph{bits per joint-sample} and as our bytes divided by ACL's on the same clips. Sizes include every header; shared model weights are reported separately.

\ifinlinefigs\input{sec/figs/fig_two_stage}\fi

\paragraph{Shell error.}
We use ACL's shell-distance metric~\cite{frechette2023acl}. Forward kinematics gives the object-space transform of every joint,
\begin{equation}
    \rotation^{o}_j = \rotation^{o}_{\parent(j)} \otimes \rotation_j, \qquad
    \textbf{x}^{o}_j = \rotation^{o}_{\parent(j)}\, \translation_j\, \overline{\rotation^{o}_{\parent(j)}} + \textbf{x}^{o}_{\parent(j)},
    \label{eq:prelim:fk}
\end{equation}
and the error of joint $j$ at sample $t$ is the largest displacement of three virtual vertices at distance $\shell = 3$\,cm along its axes,
\begin{equation}
    \err_{j,t} = \max_{a \in \{x,y,z\}} \big\| T^{o}_j(t)\,(\shell \hat{\textbf{e}}_a) - \hat{T}^{o}_j(t)\,(\shell \hat{\textbf{e}}_a) \big\|,
    \label{eq:prelim:shell}
\end{equation}
where $T^{o}_j$ and $\hat{T}^{o}_j$ are the original and decoded object-space transforms, so the error of a joint contains the errors of all its ancestors. The vertices sample the sphere, so Eq.~\eqref{eq:prelim:shell} can understate the largest displacement on it (by up to a factor $\sqrt{2/3}$ for a pure rotation error); the ACL commit we benchmark uses this approximation, which its development branch has since replaced with the exact maximum (supplementary Sec.~A). Both codecs are measured with Eq.~\eqref{eq:prelim:shell}, at the sample times only as ACL does; our implementation matches the clip maximum our ACL build reports within $10^{-4}$\,cm.

\subsection{Two-Stage Compression Pipeline}
\label{sec:prelim:pipeline}

As in \prevname{}, we view compression as two stages (Fig.~\ref{fig:prelim:pipeline}). The first, deterministic preprocessing shared by ACL and \name{}, drops the scale sub-tracks that ACL's raw convention reserves and \emph{folds} every sub-track that does not move within $p$ into a single default or constant value. Production skeletons contain many such joints: on our corpus, dropping scale reduces ACL's raw $647$\,GB to $453$\,GB, and at $p = 0.01$\,cm folding leaves only $53\,\%$ of the rotation and $3.6\,\%$ of the translation sub-tracks animated. \name{} applies ACL's folding rule unchanged, so both codecs start the second stage from the same animated sub-tracks.

\paragraph{ACL's compression stage.}
ACL drops the $w$ component of each quaternion and reconstructs it from the unit norm. It normalizes every animated sub-track to its clip range and to the range of each $16$-sample segment, and gives each joint one of $25$ bit widths, chosen per segment by a search that raises widths until the object-space error of the joint and its descendants meets $p$. We use ACL's development branch after release $2.1.0$ (commit \texttt{3ee5685}), whose newer search does not read ACL's compression level; supplementary Sec.~A describes it and compares the release. ACL performs no prediction and no entropy coding: every sample of a segment costs the same number of bits, which gives constant-time random access. It can also strip whole frames that linear interpolation of their neighbours reproduces within $p$ (on by default). Our corpus becomes an estimated $32.6$\,GB, extrapolated from the test side. \name{} replaces this stage (Sec.~\ref{sec:method}) with log-map rotations, steps and keys chosen in closed loop through forward kinematics, prediction, and learned entropy coding; the corpus then becomes an estimated $11.9$\,GB, plus one $870$\,KB model shared by all clips.

\subsection{Error Contracts}
\label{sec:prelim:gates}

ACL compresses to a precision $p$ (default $0.01$\,cm) but does not guarantee it: its widths come from a fixed set and its dropped $w$ has a precision floor, so a clip's maximum error stays within $p$ on only $1.5\,\%$ of our test clips at $0.01$\,cm. Comparing codecs at the same $p$ is fair only if both are held to the same \emph{measured} error. We check every clip against ACL's result on that clip, with $\aclmax_j$ and $\aclmean_j$ the maximum and mean error ACL reaches on joint $j$, $\aclmean$ its clip mean, and $f^{\mathrm{ACL}}$ the fraction of joint-samples it leaves above $p$.

\paragraph{Max gate.}
Every joint stays within $5\,\%$ of the larger of $p$ and ACL's own maximum on that joint, and the fraction of samples above $p$ within half a percentage point of ACL's:
\begin{equation}
\begin{aligned}
    \max_t \err_{j,t} &\leq 1.05 \max\!\big(p, \aclmax_j\big) \quad \forall j, \\
    \tfrac{1}{\nbjoint\nbframe}\big|\{(j,t) : \err_{j,t} > p\}\big| &\leq f^{\mathrm{ACL}} + 0.005.
\end{aligned}
    \label{eq:prelim:maxgate}
\end{equation}
This bounds a joint by $1.05\,p$ where ACL meets $p$ and by ACL's own maximum where it does not; a joint ACL codes far below $p$ may be coded up to $1.05\,p$. It holds ACL's worst case \emph{within tolerance}, not identically; Sec.~\ref{sec:exp:contracts} measures what the tolerances are worth.

\paragraph{Mean gate.}
The clip keeps ACL's mean $\budget = \max(\aclmean, \gfloor p)$, each joint stays near its share, and a guard bounds the worst case:
\begin{equation}
\begin{aligned}
    \tfrac{1}{\nbframe}\textstyle\sum_t \err_{j,t} &\leq \gcap \max\!\big(\budget, \aclmean_j\big) \quad \forall j, \\
    \tfrac{1}{\nbjoint\nbframe}\textstyle\sum_{j,t} \err_{j,t} &\leq \gtol\, \budget, \\
    \max_t \err_{j,t} &\leq \max\!\big(\gguard\, \budget,\; 1.05\, \aclmax_j\big) \quad \forall j,
\end{aligned}
    \label{eq:prelim:meangate}
\end{equation}
with $\gcap = 1.05$, $\gtol = 1.02$, $\gguard = 10$, and $\gfloor = 0$. The guard is ten times ACL's clip mean, not a multiple of $p$: at $p \leq 0.1$\,cm the resulting worst case is below ACL's, but at loose precision it admits large isolated errors, so the mean gate suits storage at tight precision.

\paragraph{Verification and fallback.}
The encoder decodes its own stream and checks Eq.~\eqref{eq:prelim:maxgate} or~\eqref{eq:prelim:meangate} on every joint. A failing clip is re-encoded with its internal limits scaled by $0.98$, then $0.95$, then $0.90$; if none passes, or if the final stream is larger than ACL's, ACL's own stream is stored, which satisfies either contract by definition. All test-side byte counts include this fallback.

%% file: sec/figs/fig_two_stage.tex
\begin{figure*}[t]
    \centering
    \includegraphics[width=\textwidth]{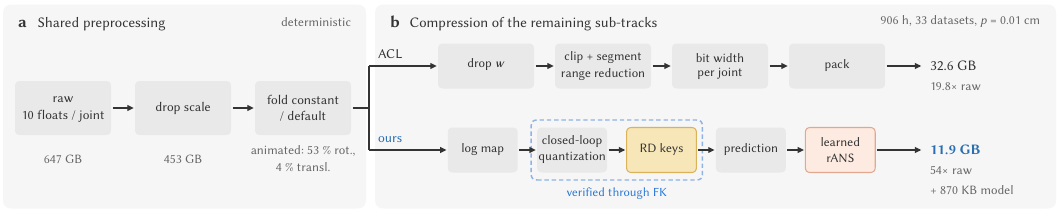}
    \caption{\capTwoStage}
    \Description{\descTwoStage}
    \label{fig:prelim:pipeline}
\end{figure*}

%% file: sec/4_method.tex
\ifinlinefigs\input{sec/figs/fig_pipeline}\fi

\section{CurveCodec 2}
\label{sec:method}

Fig.~\ref{fig:method:pipeline} gives an overview. As in \prevname{}, the unit of coding is one animated sub-track, a \emph{curve}: one joint's rotation (three log-map components) or translation (three components) over time; each coded component becomes one integer \emph{column} of the entropy coder. The skeleton enters the codec in one place: the encoder evaluates the shell error through forward kinematics to decide how coarsely each curve may be coded. The learned model never sees joint identity; its only hierarchy-derived input is a four-class depth bucket that the decoder computes from the skeleton's parent array. Sections~\ref{sec:method:repr} to~\ref{sec:method:keys} describe the lossy part, which determines the decoded clip, and Sec.~\ref{sec:method:entropy} the lossless part, which determines only the number of bits, so the network cannot violate the error contract.

\subsection{Representation}
\label{sec:method:repr}

\paragraph{Folding.}
Folding follows ACL's rule. On long chains (tails, finger rigs) small folding errors can accumulate below each joint's threshold yet push a descendant over its limit; when the unquantized state already violates the contract, we un-fold sub-tracks on the offending root paths, largest error first. In the benchmarked max-gate codec this runs only in the first quantization stage, which causes most loose-precision failures (Sec.~\ref{sec:exp:contracts}).

\paragraph{Log-map rotations.}
ACL stores the $(x,y,z)$ part of a quaternion and reconstructs $w$, which amplifies quantization error near $|w| \approx 0$ and leaves a precision floor: on our curated set ACL's maximum error is $0.071$\,cm at every precision from $0.01$ down to $0.001$\,cm. We instead code the rotation vector~\cite{grassia1998practical}
\begin{equation}
    \logrot = \Log(\rotation) = 2\,\mathrm{atan2}\big(\norm{\rotation_{xyz}}, q_w\big)\, \frac{\rotation_{xyz}}{\norm{\rotation_{xyz}}},
    \label{eq:method:logmap}
\end{equation}
after choosing quaternion signs so that consecutive samples lie in the same hemisphere ($\rotation_t \cdot \rotation_{t-1} \geq 0$, with $q_w \geq 0$ at $t = 0$), so that $\norm{\logrot} \in [0, 2\pi]$. Where $\norm{\rotation_{xyz}}$ vanishes (the identity, or $\rotation = -1$ after a full turn) the direction is undefined and the limit depends on the direction of approach; by convention we take $\logrot = 0$, a branch choice that $\Exp$ maps back to the same rotation in both cases. The map needs no side information, bounds the rotation error by $\norm{\Delta\logrot}$, and lowers the curated maximum at $0.01$\,cm from $0.071$ to $0.024$\,cm. It is continuous only short of a full turn: when a joint's rotation relative to its parent passes $2\pi$, $\logrot$ jumps from about $+2\pi\,\textbf{u}$ to $-2\pi\,\textbf{u}$. Since $\logrot$ and $\logrot + 2\pi\,\textbf{u}$ are the same rotation in $SO(3)$ (their quaternions differ only in sign) and the contract is checked on the decoded clip, this costs a few keys, not correctness; it occurs on $0.33\,\%$ of the test side's rotation tracks and costs below $0.2\,\%$ of the bytes (supplementary Sec.~C, which also evaluates an unwrapping rule). Translations are coded in centimetres without any transform.

\subsection{Closed-Loop Quantization}
\label{sec:method:quant}

Each curve is quantized uniformly, $\qint_t = \rint(v_t / \step)$, $\hat v_t = \qint_t \step$, with the step on a logarithmic grid $\step = 2^{\gidx / \gridres}$, $\gidx \in \integers$, $\gridres = 16$, so that the header stores a small integer. Because a joint's error is amplified by the chain it moves, we start from
\begin{equation}
    \step_{\mathrm{rot}} = \frac{2\, p}{\sqrt{3}\, \rdom_j \sqrt{\chain_j}}, \qquad
    \step_{\mathrm{trans}} = \frac{p}{\sqrt{3}\, \sqrt{\chain_j}},
    \label{eq:method:init}
\end{equation}
where $\rdom_j$ is the shell distance plus joint $j$'s longest child chain and $\chain_j$ the number of joints on the longest root-to-leaf chain through $j$. A closed loop then grows each track's step while the contract holds: every candidate is decoded, forward kinematics is recomputed for the joint's subtree, and Eq.~\eqref{eq:prelim:maxgate} or~\eqref{eq:prelim:meangate} is tested on every joint of the subtree. The search passes leaves to root and back, offering $1, 2, 4, \ldots$ grid units and bisecting on refusal; because the error is not monotone in the step, it looks up to half an octave past a refusal, which alone saves $1$ to $7\,\%$ under the mean gate. Under the max gate the three rotation components may then grow separately. The root keeps the finest steps; leaves end $2$ to $5\times$ coarser than Eq.~\eqref{eq:method:init}.

\subsection{Rate--Distortion Keys}
\label{sec:method:keys}

Each track keeps a key set $\keys \subseteq \{0, \ldots, \nbframe-1\}$ with the first and last sample, and the decoder fills the rest by Catmull--Rom interpolation~\cite{catmull1974splines} of the log-map values,
\begin{equation}
\begin{aligned}
    \hat{\logrot}(t) &= h_{00}(u)\,\hat{\logrot}_{k} + h_{10}(u)\,\Delta_k \textbf{m}_{k} + h_{01}(u)\,\hat{\logrot}_{k'} + h_{11}(u)\,\Delta_k \textbf{m}_{k'}, \\
    \textbf{m}_{k} &= \frac{\hat{\logrot}_{k^+} - \hat{\logrot}_{k^-}}{t_{k^+} - t_{k^-}}, \qquad u = \frac{t - t_k}{\Delta_k},
\end{aligned}
    \label{eq:method:hermite}
\end{equation}
with consecutive keys $k, k'$, $\Delta_k = t_{k'} - t_k$, neighbouring keys $k^\pm$, and cubic Hermite basis functions $h_{\cdot\cdot}$; rotations are recovered with $\Exp$. The cubic saves $0.5$ to $2.6\,\%$ over linear interpolation in the same rotation-vector coordinates; this is not spherical linear interpolation~\cite{shoemake1985animating}, which it matches only for coaxial rotations, and every interpolation baseline in this paper is log-linear. Keys, like steps, are chosen in closed loop through the hierarchy.

\paragraph{Max gate.}
From the all-samples state, the encoder removes keys in greedy rounds: it proposes every fourth interior key of every track, decodes the affected window, recomputes the subtree's error, and accepts the removal if every joint stays within its limit and the clip within its over-$p$ allowance. A narrow failure may be rescued by moving the two neighbouring key values up to two steps towards their least-squares fit of the window. Rounds alternate leaf-to-root and root-to-leaf passes until nothing is accepted.

\paragraph{Mean gate.}
Under the iso-mean contract the budget is a sum, so moves can trade between curves. A Lagrangian ladder~\cite{sullivan1998rd} lets two moves compete: removing a key, with distortion $\dist$ the increase of the summed error over the window and subtree and rate $\rate$ estimated from the residuals it removes, and growing a step by $g$ grid units, with $\rate = \gbits \cdot 3\, |\keys|\, g / \gridres$ and $\gbits = 2.5$ calibrated against measured bits. At level $\lagr$ every move with $\dist / \rate \leq \lagr$ that keeps Eq.~\eqref{eq:prelim:meangate} is accepted; $\lagr$ starts at $10^{-4}\budget$ per bit and grows by $\sqrt{2}$ until the budget is spent, about $60\,\%$ on steps and $40\,\%$ on keys.

\paragraph{Track modes.}
Each track is finally coded in the cheapest of four modes: static, all samples, one key set shared by its components, or per-component sets that may only \emph{thin} the shared set, with a context-coded presence flag per component and key.

\begin{algorithm}[t]
\SetAlgoLined
\small
\KwIn{clip $\motion$, precision $p$, contract (max or mean), ACL's per-joint statistics}
\KwOut{key sets $\keys$, step indices $\gidx$, integers $\qint$}
fold constant and default sub-tracks (ACL's rule); un-fold on chains that break the contract\;
convert rotations to log maps with sign continuity along time (Eq.~\ref{eq:method:logmap})\;
initialize steps (Eq.~\ref{eq:method:init}); $\keys \leftarrow$ all samples\;
\Repeat(\tcp*[f]{object-space step search}){no step changes}{
  \For{joints $j$ leaf-to-root, then root-to-leaf}{
    grow $\gidx_j$ by $1,2,4,\ldots$ grid units with lookahead; accept if the subtree satisfies the contract\;
  }
}
\eIf{max gate}{
  \Repeat{no key removed}{remove candidate keys whose subtree stays within its limits (with neighbour refit)\;}
}{
  \For{$\lagr \leftarrow 10^{-4}\budget,\ \sqrt{2}\lagr,\ \ldots$}{
    accept key removals and step growths with $\dist/\rate \leq \lagr$ that keep Eq.~\eqref{eq:prelim:meangate}\;
  }
}
choose per-track mode and component key thinning by coded cost\;
\caption{Lossy encoder of \name{}. Every acceptance test decodes the candidate and evaluates the shell error of the affected subtree exactly.}
\label{alg:method:encoder}
\end{algorithm}

\subsection{Prediction and Learned Entropy Coding}
\label{sec:method:entropy}

After the lossy stage a clip is a set of integer columns, the quantized values of each coded component at its keys (at most six per joint, $468$ for a $78$-joint clip before folding).

\paragraph{Prediction and alphabet.}
Each column is predicted from its own past, $\resid_t = \qint_t - \pred_t$, with a predictor chosen per column among first and second differences, a second difference scaled by the ratio of key gaps, and an order-$4$ integer NLMS filter~\cite{haykin2002adaptive} on first differences. A residual is split into a symbol (zero or the sign, the bit length of $|\resid_t|$, and the first bit below its leading one; $159$ symbols) and raw low bits. Each symbol covers an interval of residual values, so a continuous distribution over $\resid_t$ induces symbol probabilities through its cumulative distribution.

\paragraph{Learned probability model.}
A network $\entnet$ sees only the column's own decoded past through $36$ integer features $\feat_t$: the last eight residuals as signed log magnitudes, five exponential averages of $|\resid|$, first and second differences of the decoded integers, the key gaps around the position, the depth class (root, one to two, three to five, six and deeper), the sub-track kind, predictor, log step, frame rate, precision, and position in the column; none identifies a joint, a skeleton, or a dataset. $\entnet$ is a causal transformer, two pre-norm blocks of width $64$ with four heads over the last $\ctxwin = 128$ positions ($108$\,K parameters), whose head outputs a mixture of $\nbmix = 3$ logistics~\cite{salimans2017pixelcnnpp}; the symbol covering $[a, b]$ has probability
\begin{equation}
    P(a \leq \resid_t \leq b \mid \feat_{\leq t}) = \sum\nolimits_{k=1}^{\nbmix} w_k \big[ S_k(b + \tfrac12) - S_k(a - \tfrac12) \big],
    \label{eq:method:mixture}
\end{equation}
with $S_k(x) = \mathrm{sig}\big((x - \mu_k)/\sigma_k\big)$, quantized to $15$-bit frequencies and coded with rANS~\cite{duda2013ans}. Residuals are ordered time-major over all columns, so the model runs once per index as a batch over the active columns, and the decoder reconstructs each integer before the next features are computed. For the max-gate codec we use a $22$\,K-parameter MLP ($36 \to 128 \to 128 \to 9$) over the same features, $4\times$ cheaper to decode.

\paragraph{Bit-exact inference.}
Encoder and decoder must compute identical probabilities. We export the network to fixed point: $16$-bit weights with power-of-two scales, $64$-bit integer accumulators, integer layer normalization, and lookup tables for softmax, sigmoid, and exponential. The residual stream is clipped to $\pm 2^{28}$, which bounds every operand downstream: linear-layer products stay below $2^{43}$ and their sums over at most $256$ terms below $2^{53}$, so even a float64 matrix library adds exact integers~\cite{balle2019integer}; the attention logits and the layer-norm variance are accumulated in $64$-bit integers within the bounds given in Appendix~H. The guarantee is that every probability and every decoded integer is identical on any platform. The float reconstruction (dequantization, interpolation, $\Exp$) runs on those integers and reproduced x86 results bit for bit on an ARM machine in our tests, but it is deterministic rather than exact by construction (Appendix~H).

\paragraph{Training.}
The model minimizes the code length of Eq.~\eqref{eq:method:mixture} on residual streams of our training split ($886$\,h, $41.2$\,G symbols), on $40$\,M windows with $128$ context positions and $384$ scored ones, with AdamW~\cite{loshchilov2019adamw} and a one-cycle schedule, in about $1.5$ hours on one RTX 4090. Supplementary Appendix~G summarizes the bitstream layout; the reference decoder defines the details it leaves out.

%% file: sec/figs/fig_pipeline.tex
\begin{figure*}[t]
    \centering
    \includegraphics[width=\textwidth]{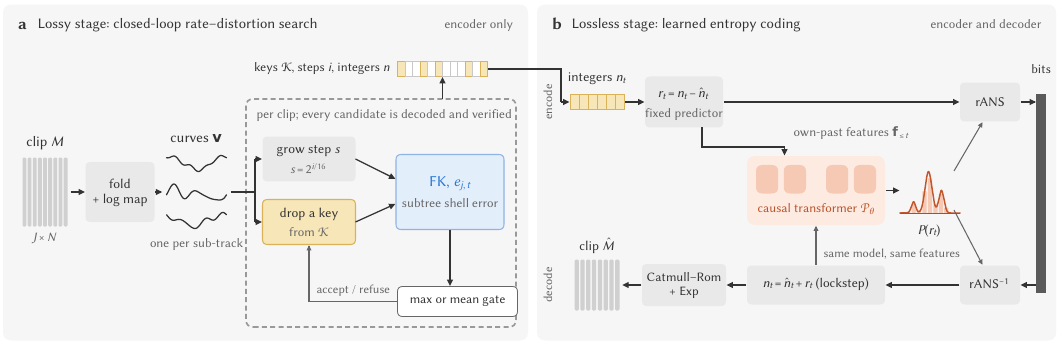}
    \caption{\capPipeline}
    \Description{\descPipeline}
    \label{fig:method:pipeline}
\end{figure*}

%% file: sec/5_analysis.tex
\section{Where Is the Redundancy in Motion?}
\label{sec:analysis}

These measurements motivate the design. Unless stated otherwise, they use two development sets that predate the full corpus: \emph{curated}, $70$ clips from $14$ mocap datasets ($15.9$\,M joint-samples), and \emph{dev}, $15$ clips. Every variant was verified against its contract on every clip; Sec.~\ref{sec:exp} tests the final codec on held-out data.

\subsection{ACL's Bit Budget}
\label{sec:analysis:acl}

\ifinlinefigs\input{sec/figs/fig_acl_budget}\fi
Fig.~\ref{fig:analysis:acl}a follows ACL's stages on the curated set. Two stages do the work: folding sub-tracks that do not move ($2.84\times$, from $191$ to $68$ bits per joint-sample) and range reduction with variable bit widths ($3.74\times$, to $18.1$ bits); the $4\times$ spread of ACL's ratio across datasets comes from folding. Of the remaining bytes, $81\,\%$ are animated rotations and $13.5\,\%$ per-segment range metadata, a share that grows to $24\,\%$ at $0.1$\,cm on the test side because it does not shrink with the precision. Fig.~\ref{fig:analysis:acl}b asks what a lossless back end could add. The zeroth-order entropy of ACL's quantized samples is $97.7\,\%$ of the bits it spends, so its widths are nearly tight for independent samples; first and second differences reach $83.8$ and $81.1\,\%$, and real context coders $0.80\times$. These bounds hold for the models we tried, up to second-order temporal context; little is left because range reduction stretches every $16$ samples to fill their integer range.

\emph{Consequence.} A substantially smaller stream needs its own quantizer. Ours is close to ACL in zeroth-order entropy ($0.92$ to $1.07\times$), but its first differences cost $0.57$ to $0.62\times$ ACL's bits against $0.84\times$ for ACL's own symbols: the gain appears once time is modelled.

\subsection{Samples Predicted and Samples Not Coded}
\label{sec:analysis:keys}

\ifinlinefigs\input{sec/figs/fig_keys}\fi
\paragraph{Temporal predictability.}
Predicting each integer from its own past reduces the log-map stream by $39$ to $55\,\%$ against its zeroth-order cost. The best predictor depends on the joint and capture: second differences beat first differences by $18\,\%$ at the root but only $4$ to $7\,\%$ at depth six and below, where noise at the scale of the step dominates, and an adaptive (NLMS) predictor wins on $37$ to $44\,\%$ of the curves, which motivates a per-curve choice.

\paragraph{Per-joint sparsity.}
Not every sample needs coding. Under the max gate the encoder keeps $87\,\%$ of the root's samples but only $57\,\%$ at depth six and below at $0.01$\,cm, and $51$ against $22\,\%$ at $0.1$\,cm (Fig.~\ref{fig:analysis:keys}a; five development clips, an earlier version of the encoder); the root keeps the most because its error reaches every descendant. ACL can only strip whole frames whose every joint is linearly interpolable, $0.38\,\%$ of the curated frames at $0.01$\,cm. Where keys pay, the gain comes from the closed loop: removing keys by a local per-curve error bound saves only $0$ to $0.6\,\%$.

\paragraph{The price of a bit.}
At the margin, keys are the cheapest way to spend error. Fig.~\ref{fig:analysis:keys}c compares the distortion the iso-mean encoder adds per bit it saves at $0.1$\,cm: growing a step by one grid unit costs $0.17$ step units, a dead zone $0.31$, and removing an RD-selected key about $0.03$. This says which move to prefer, not how much each move contributes in total, which depends on how many cheap moves the precision leaves: on held-out data, removing keys from the final codec costs $8\,\%$ at $0.01$\,cm and $69\,\%$ at $1$\,cm, where it matches prediction (Sec.~\ref{sec:exp:ablation}). Codecs that code every sample lose accordingly: a dyadic hierarchical codec with a learned in-betweener matches ours where keys keep $83\,\%$ of the samples ($1.01\times$) but needs $1.55$ to $1.78\times$ the bytes where they keep $7$ to $15\,\%$.

\emph{Consequence.} Prediction is the largest saving at $p \leq 0.1$\,cm and the choice of which samples to omit the second, with a share that grows as $p$ loosens. The encoder needs both, with steps and keys competing for one budget.

\subsection{Interpolation Gains Little on the Gaps the Encoder Leaves}
\label{sec:analysis:interp}

\ifinlinefigs\input{sec/figs/fig_oracle}\fi
Where keys carry the bytes, a better interpolator should buy more: one that halved the cubic error everywhere ($k = 2$) would save $3$ to $27\,\%$. This is the bet of \prevname{} and of learned in-betweening; we tested it, and the results below bound what the estimators we tried achieved on our data, not what any interpolator could. The gaps the encoder leaves are short: under the max gate at $p \leq 0.1$\,cm, $77$ to $92\,\%$ hide at most two samples (Fig.~\ref{fig:analysis:keys}b), whereas published in-betweeners train on gaps of $5$ to $45$ frames at errors $60\times$ above our default precision. In our gaps the cubic residual is linearly unpredictable from the curve's own context (ridge $R^2 \leq 0.02$).

\paragraph{A retrieval oracle.}
As a strong nonparametric reference, an oracle searches every window of $510$ training clips ($4.2$\,M samples, held-out skeletons) for the best match of a gap's decoded context and copies the matched interior as a residual over cubic interpolation. It is a nearest neighbour in context space over those $4.2$\,M samples, a $510$-clip subset of the $886$-hour training split the entropy model sees. On random gaps it beats cubic interpolation by $1.37$, $1.29$, and $1.23\times$ at $p = 0.1$, $0.3$, and $1$\,cm and keeps improving with the corpus size (Fig.~\ref{fig:analysis:oracle}). On the gaps the iso-mean RD encoder leaves, it reaches only $0.85$ to $0.90$, no better than \emph{linear} interpolation on the same gaps, and is flat over the $32$-fold range of library size we measured. The encoder removes exactly the samples a smooth prior already predicts and keeps contacts, impacts, and direction changes with no precursor in the curve's past. A better interpolator might also let the encoder drop keys it now keeps; we placed four learned in-betweeners inside the encoder, including a $3.4$\,M-parameter transformer with keys re-chosen for it. None paid (supplementary Sec.~F): they changed the size by $-2.5$ to $+5.7\,\%$, no more than the null option of choosing per gap between cubic and linear interpolation would give.

\emph{Consequence.} Under an explicit error contract and an explicit bit count, none of the interpolators we built found enough information in the transmitted context to pay for itself. These experiments ran under the iso-mean contract; for the max gate, whose gaps are shorter still, the conclusion is an inference we did not test. \name{} therefore keeps a fixed interpolator and moves the learned prior to the probability model.

\subsection{Where a Network Pays}
\label{sec:analysis:net}

\begin{table}[t]
    \centering
    \caption{Every learned role we tested, each as a complete codec under the iso-mean contract and verified on every clip (dev set; bytes relative to our codec with the MLP entropy model). Only the learned probability model of the residuals pays.}
    \label{tab:analysis:neural}
    \small
    \setlength{\tabcolsep}{3pt}
    \begin{tabular}{llccc}
        \toprule
        Network role & Representation & $0.01$ & $0.1$ & $1$\,cm \\
        \midrule
        Entropy model & keys + residuals & \textbf{0.984} & \textbf{0.982} & \textbf{0.983} \\
        Correction & keys + fixes & \multicolumn{3}{c}{$1.00$ to $1.04$ (unused)} \\
        Transform + model & lifting wavelet & 1.03 & 1.09 & 1.15 \\
        In-betweener + model & dyadic grid & 1.10 & 1.35 & 1.94 \\
        Tokenizer & residual VQ & 1.10 & 1.39 & 2.09 \\
        Neural field & per-window INR & -- & 2.0 & 3.3 \\
        \bottomrule
    \end{tabular}
\end{table}

Tab.~\ref{tab:analysis:neural} summarizes every role we gave a network. Dense learned representations have a precision floor: a residual-VQ tokenizer cannot represent a track below about $1.3\,\%$ of its RMS and a meta-learned neural field not below $1.6$ to $4.8\,\%$ of the window's amplitude, while the contract asks for $0.01$ to $0.1\,\%$, so both degenerate into scalar residual codecs. In a lifting-wavelet codec of ours, learning the transform changed the size by $\pm 1\,\%$ while learning its entropy model saved $7$ to $12\,\%$; learned image codecs report a similar division of gains~\cite{balle2018variational,minnen2018joint}, an analogy rather than evidence for our numbers. Over a hand-designed context coder, the MLP entropy model saves $3$ to $7\,\%$ on mocap and $13\,\%$ on game assets. It does not predict the next residual's value ($R^2 \leq 0.02$) but its distribution, from a slowly varying per-curve regime: the residual's scale relative to the step, periodicity, and whether a curve is held, linear, or noisy. Eight positions of context give $60$ to $70\,\%$ of the gain. Supplementary Sec.~F reports that cross-joint structure (mirroring, pose-space PCA, interpolation context) and transforms under a worst-case contract did not pay on our data.

\emph{Consequence.} The network belongs in the probability model, where it models a distribution the codec must send anyway, and it can be small, causal, and local.

%% file: sec/figs/fig_acl_budget.tex
\begin{figure}[t]
    \centering
    \includegraphics[width=\columnwidth]{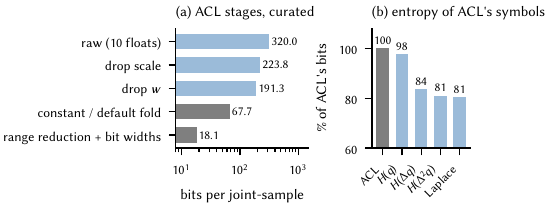}
    \caption{\capAclBudget}
    \Description{\descAclBudget}
    \label{fig:analysis:acl}
\end{figure}

%% file: sec/figs/fig_keys.tex
\begin{figure*}[t]
    \centering
    \includegraphics[width=\textwidth]{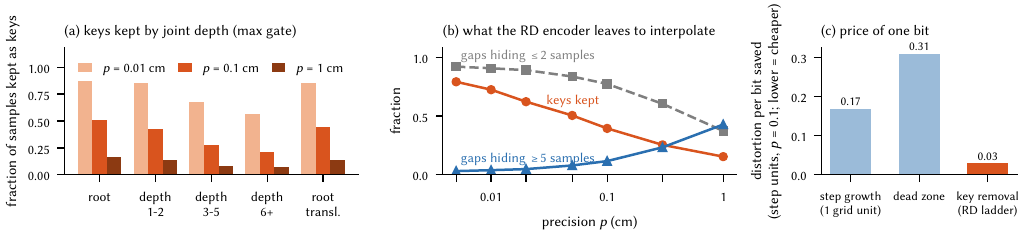}
    \caption{\capKeys}
    \Description{\descKeys}
    \label{fig:analysis:keys}
\end{figure*}

%% file: sec/figs/fig_oracle.tex
\begin{figure}[t]
    \centering
    \includegraphics[width=\columnwidth]{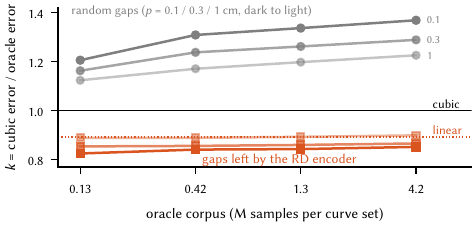}
    \caption{\capOracle}
    \Description{\descOracle}
    \label{fig:analysis:oracle}
\end{figure}

%% file: sec/6_exp.tex
\section{Experiments}
\label{sec:exp}

\subsection{Setup}
\label{sec:exp:setup}

\paragraph{Corpus and split.}
We assembled $192{,}392$ clips ($906$\,h, $16.17$\,G joint-samples) from $33$ datasets, including HiPHI~\cite{ji2026hiphi}, a large mocap dataset released online (\emph{large online mocap} below), Motion-X~\cite{lin2023motionx}, BEAT~\cite{liu2022beat}, AMASS~\cite{mahmood2019amass}, MotionPersona~\cite{shi2025motionpersona}, 100STYLE~\cite{style100}, CMU~\cite{CMU:mocap}, LAFAN1~\cite{harvey2020robust}, ZeroEGGS~\cite{ghorbani2023zeroeggs}, and PFNN~\cite{holden2017pfnn}, retargeted and converted versions of several of them, hand capture, and $549$ game-asset clips of $196$ skeletons, at $20$ to $120$\,Hz. Because the corpus holds retargeted, cropped, and converted versions of the same performances, we split by \emph{take group}, a key that follows the original performance across datasets and conversions, holding out $2$ of every $100$ clips per dataset: the test side has $4{,}472$ clips, $20.1$\,h, and $342.7$\,M joint-samples, and no take group appears on both sides. A quadruped dataset~\cite{zhang2018mode} (dog) is held out of training entirely. The test side also contains the $71$ development clips of Sec.~\ref{sec:analysis} ($5.1\,\%$ of its joint-samples); without them the ratios change by less than one point.

\paragraph{Baselines and metrics.}
ACL~\cite{frechette2023acl} (commit \texttt{3ee5685}) runs with its default settings at $p = 0.005$, $0.01$, $0.02$, $0.05$, $0.1$, $0.3$, and $1$\,cm, with keyframe stripping off because our contracts take their targets from the no-strip result; we report our ratio against ACL's default stripping as well. \name{} runs under the max gate with the MLP entropy model and under the mean gate with the transformer, our product configuration. Learned motion encoders are not included; their errors are two orders of magnitude above ACL's at these precisions~\cite{shi2026curvecodec}. We report bits per joint-sample, the ratio of our bytes to ACL's on the same clips, and the mean, $99$th-percentile, and maximum shell error over all joint-samples. Timing uses one pinned core of an AMD Threadripper PRO 7975WX for both codecs.

\subsection{Compression against ACL}
\label{sec:exp:rd}

\ifinlinefigs\input{sec/figs/fig_rd}\fi

\begin{table*}[t]
    \centering
    \caption{Compression and error on the held-out test side ($4{,}472$ clips, $342.7$\,M joint-samples). Ratio is our bytes divided by ACL's, after the fallback (Sec.~\ref{sec:prelim:gates}); errors are shell errors in cm over all joint-samples of the delivered streams. The max gate holds every joint within $5\,\%$ of the larger of $p$ and ACL's maximum and spends the room below the worst case, so its mean error is higher; the mean gate holds ACL's mean with a flatter distribution (higher P99) and a lower maximum for $p \leq 0.1$\,cm. At $0.3$ and $1$\,cm ACL misses its own precision on $6.7$ and $13.5\,\%$ of its joint-samples, and the mean gate's guard admits the large maxima shown; ratios in parentheses are on the clips where ACL keeps at least $99\,\%$ of its joint-samples within $p$.}
    \label{tab:exp:main}
    \small
    \setlength{\tabcolsep}{4pt}
    \begin{tabular}{l cccc ccccc ccccc}
        \toprule
        & \multicolumn{4}{c}{ACL} & \multicolumn{5}{c}{Ours, max gate} & \multicolumn{5}{c}{Ours, mean gate} \\
        \cmidrule(lr){2-5}\cmidrule(lr){6-10}\cmidrule(lr){11-15}
        $p$ (cm) & bits & Mean & P99 & Max & ratio & bits & Mean & P99 & Max & ratio & bits & Mean & P99 & Max \\
        \midrule
        0.005 & 18.43 & -- & -- & -- & 0.414 & 7.62 & -- & -- & -- & & & & & \\
        0.01 & 16.38 & 0.0034 & 0.0111 & 0.213 & \textbf{0.373} & 6.11 & 0.0048 & 0.0120 & 0.211 & \textbf{0.373} & 6.11 & 0.0033 & 0.0119 & 0.094 \\
        0.02 & 14.39 & -- & -- & -- & 0.325 & 4.68 & -- & -- & -- & & & & & \\
        0.05 & 11.87 & 0.0166 & 0.0508 & 0.539 & \textbf{0.262} & 3.11 & 0.0236 & 0.0540 & 0.536 & 0.267 & 3.17 & 0.0166 & 0.0576 & 0.480 \\
        0.1 & 10.11 & 0.0333 & 0.1003 & 0.916 & \textbf{0.217} & 2.19 & 0.0466 & 0.1054 & 0.845 & 0.220 & 2.23 & 0.0332 & 0.1155 & 0.903 \\
        0.3 & 8.53 & 0.137 & 0.379 & 2.39 & 0.174 (0.159) & 1.48 & 0.173 & 0.395 & 2.38 & 0.137 (0.157) & 1.16 & 0.134 & 0.445 & 7.9 \\
        1 & 8.72 & 0.560 & 1.47 & 7.17 & 0.148 (0.136) & 1.29 & 0.646 & 1.50 & 6.97 & 0.069 (0.137) & 0.60 & 0.542 & 1.73 & 24.7 \\
        \bottomrule
    \end{tabular}
\end{table*}

At ACL's default precision of $0.01$\,cm, \name{} needs $6.1$ bits per joint-sample where ACL needs $16.4$, $0.37\times$ ACL's bytes under either contract (Tab.~\ref{tab:exp:main}, Fig.~\ref{fig:exp:rd}). The ratio improves as $p$ loosens because ACL's segment metadata and widths have a floor while our keys thin out. Under the max gate our P99 is within $1.08\times$ of ACL's and our mean error $1.4\times$ higher; under the mean gate our mean matches ACL's on every clip and our maximum is lower at $p \leq 0.1$\,cm ($0.094$ against $0.213$\,cm at $0.01$\,cm). At $0.3$ and $1$\,cm ACL itself exceeds $p$ on part of the data, which both contracts inherit. Against ACL with its default keyframe stripping, which removes $4.8\,\%$ of the frames at $0.01$\,cm and $29\,\%$ at $0.1$\,cm without changing its error, our ratios are $0.380\times$ and $0.25$ to $0.26\times$. ACL's compression level changes no byte in our build. The released ACL $2.1.0$ writes $2$ to $7\,\%$ more bytes at a tighter worst case, a comparison of bytes only, since our contracts were not re-matched to its errors. xz or zstd over ACL's stream saves $2$ to $5\,\%$ (supplementary Sec.~A). Extrapolated per dataset (Fig.~\ref{fig:exp:storage}), our $906$-hour corpus would take an estimated $11.9$\,GB at $0.01$\,cm, against the $32.6$\,GB that ACL would need.

\ifinlinefigs\input{sec/figs/fig_storage}\fi
\ifinlinefigs\input{sec/figs/fig_qualitative}\fi
Fig.~\ref{fig:exp:qualitative} shows decoded motion for a humanoid, a game character, a creature, and the held-out dog. At $0.01$\,cm, decoded and original skeletons coincide for all three codecs while our streams are $2$ to $3\times$ smaller; at $1$\,cm the largest joint displacement in the frames shown is about half a centimetre for all three (the clip maxima are in the caption), at $4.4$ to $4.7\times$ fewer bytes than ACL. A static frame cannot show popping; the evidence about the worst frame is the contracts, checked on every sample.

\subsection{Contracts: Tolerances and Failures}
\label{sec:exp:contracts}

The tolerances of the max gate are a convenience for the search, not a source of the gain. Removing them costs $0.7$ to $1.0\,\%$ of the bytes on a $311$-clip test subset, and a standalone contract that holds every joint-sample within $p$, which ACL itself meets on only $2\,\%$ of the clips, still gives $0.438\times$ and $0.265\times$ ACL's bytes at $0.01$ and $0.1$\,cm: ACL's own overshoot of $p$ is worth $4$ to $6\,\%$ of our bytes. The fallback is a safety net, not a source of compression: at $p \leq 0.1$\,cm the first encode passes on at least $99.6\,\%$ of the $4{,}472$ clips under either contract, and before any fallback no stream is larger than ACL's. The final check fails on $16$ and $3$ clips at $0.01$ and $0.1$\,cm under the max gate, nearly all hairline misses of the over-$p$ fraction, and on $4$ and $3$ under the mean gate; a margin re-encode or ACL's stream resolves every one and changes no ratio by more than $0.001$; the re-encodes of all failing clips cost about $7$ core-hours in total. At $0.3$ and $1$\,cm, $34$ and $95$ max-gate clips fail because the encoder's late step refinement cannot un-fold a folded track that already exceeds the contract, and $36$ and $92$ streams, nearly all of them these clips, are larger than ACL's before the fallback; after it, ACL's stream is stored for $13$ and $29$ clips, the counts and ratios of Tab.~\ref{tab:exp:main} (details and an un-folding fix in supplementary Sec.~B). Under the mean gate no stream is larger than ACL's at any precision.

\subsection{Generalization}
\label{sec:exp:general}

\begin{table}[t]
    \centering
    \caption{Ratio to ACL by data type and frame rate (test side, mean gate). The gain holds on every type, including game assets and a species absent from training, and grows with the frame rate.}
    \label{tab:exp:types}
    \small
    \setlength{\tabcolsep}{4pt}
    \begin{tabular}{lccc}
        \toprule
        Data type & $p = 0.01$ & $0.1$ & $1$\,cm \\
        \midrule
        native-skeleton mocap & 0.358 & 0.202 & 0.055 \\
        retargeted mocap & 0.489 & 0.322 & 0.168 \\
        converted SMPL & 0.413 & 0.325 & 0.211 \\
        game assets & 0.353 & 0.297 & 0.219 \\
        hands & 0.373 & 0.239 & 0.129 \\
        dog (held-out species) & 0.532 & 0.350 & 0.168 \\
        \midrule
        Frame rate (Hz) & \multicolumn{3}{c}{$20$ / $24$--$30$ / $60$ / $90$ / $120$ at $p = 0.1$} \\
        & \multicolumn{3}{c}{0.323 / 0.341 / 0.316 / 0.200 / 0.178} \\
        \bottomrule
    \end{tabular}
\end{table}

\ifinlinefigs\input{sec/figs/fig_per_dataset}\fi
Every dataset is smaller than ACL at every precision (Fig.~\ref{fig:exp:dataset}, Tab.~\ref{tab:exp:types}). The gain is largest on long, high-rate captures (large online mocap $0.236\times$ at $0.01$\,cm) because smoother curves let a fixed budget buy longer key gaps, while ACL's rate per sample barely changes with the frame rate. A looser bound pays only where there are samples to omit: from $0.01$ to $1$\,cm the ratio of $120$\,Hz BEAT falls $27$-fold ($0.399$ to $0.015\times$) and that of $20$\,Hz HumanAct12 $1.5$-fold ($0.676$ to $0.444\times$); the frame-rate classes are partly confounded with datasets. Retargeted data gain less because ACL is already efficient on them. The dog, a species absent from training, reaches $0.532\times$, $0.350\times$, and $0.168\times$. The entropy model is the only trained part: on the dog it saves $5.7$ to $6.9\,\%$ over the codec without a network, $91$ to $95\,\%$ of what a model trained on dog data saves. A model trained on one $120$\,Hz dataset alone is $1.34$ to $1.43\times$ \emph{larger} than no model on the dog, so the product model is trained on the whole corpus.

\subsection{Ablation}
\label{sec:exp:ablation}

\begin{table}[t]
    \centering
    \caption{Ablation of the final codecs on held-out data ($311$-clip test subset, all $33$ datasets). Top of each block: components added one at a time, bytes relative to ACL. Bottom: one component removed from the final encoder, bytes relative to the final encoder. A failing clip counts at ACL's bytes. At $1$\,cm the max-gate rows include five clips of the encoder weakness of Sec.~\ref{sec:exp:contracts}; without them the final codec reaches $0.110$. Prediction is measured on the every-sample codec.}
    \label{tab:exp:ablation_test}
    \small
    \setlength{\tabcolsep}{4pt}
    \resizebox{\columnwidth}{!}{%
    \begin{tabular}{lcccc}
        \toprule
        & $p = 0.01$ & $0.1$ & $0.3$ & $1$\,cm \\
        \midrule
        \multicolumn{5}{l}{\emph{Max gate, bytes / ACL}} \\
        log-map quantizer, every sample & 0.978 & 0.860 & -- & 0.596 \\
        + prediction and context rANS & 0.480 & 0.394 & 0.406 & 0.310 \\
        + fine step grid, lookahead, final container & 0.458 & 0.366 & 0.372 & 0.307 \\
        + RD keys & 0.432 & 0.274 & 0.265 & 0.191 \\
        + component key thinning & 0.430 & 0.271 & -- & 0.187 \\
        + MLP entropy model (final) & \textbf{0.413} & \textbf{0.256} & \textbf{0.251} & \textbf{0.182} \\
        \multicolumn{5}{l}{\emph{Max gate, final without one component, bytes / final}} \\
        without prediction (bytes / every-sample codec) & 2.34 & 2.95 & 2.77 & 2.99 \\
        without keys & 1.076 & 1.394 & 1.46 & 1.69 \\
        without MLP entropy model & 1.040 & 1.062 & -- & 1.028 \\
        without thinning & 1.003 & 1.002 & -- & 1.012 \\
        \midrule
        \multicolumn{5}{l}{\emph{Mean gate, bytes / ACL}} \\
        RD ladder, no model, no thinning & 0.456 & 0.295 & -- & 0.126 \\
        + component key thinning & 0.455 & 0.287 & -- & 0.122 \\
        + MLP entropy model & 0.440 & 0.272 & -- & 0.118 \\
        MLP replaced by transformer (final) & \textbf{0.418} & \textbf{0.260} & \textbf{0.169} & \textbf{0.113} \\
        \multicolumn{5}{l}{\emph{Mean gate, final without one component, bytes / final}} \\
        without keys & 1.317 & 1.813 & 2.207 & 2.23 \\
        without entropy model & 1.090 & 1.104 & -- & 1.074 \\
        transformer replaced by MLP & 1.052 & 1.048 & -- & 1.039 \\
        without thinning & 0.994 & 1.006 & -- & 1.016 \\
        \bottomrule
    \end{tabular}}
\end{table}

Tab.~\ref{tab:exp:ablation_test} adds components one at a time and removes each from the final encoders. Prediction is the largest component at $p \leq 0.1$\,cm: against the quantized integers coded with a general-purpose compressor (first row) it is worth $2.0$ and $2.2\times$, and $2.3$ to $3.0\times$ against the same context coder without prediction. Keys are second and grow as $p$ loosens: removing them costs $8\,\%$, $39\,\%$, and $69\,\%$ at $0.01$, $0.1$, and $1$\,cm under the max gate ($2.2\times$ at $1$\,cm without the five weakness clips) and $32\,\%$ to $2.2\times$ under the mean gate, so at $1$\,cm keys and prediction are comparable ($1.9\times$). Each configuration spends a different share of the error budget, so the table ranks contributions under the contract as used, not at strictly equal distortion. The final step search and container save $1$ to $8\,\%$; the entropy model saves $3$ to $6\,\%$ under the max gate and $7$ to $10\,\%$ under the mean gate, of which the transformer contributes $4$ to $5\,\%$ over the MLP; thinning saves at most $1.6\,\%$. Supplementary Sec.~D gives the development ablation on the curated set and the capacity of the entropy model: deeper models save another $1$ to $2.7$ points at $1.8$ to $27.5\times$ the decode cost.

\subsection{Runtime and Deployment}
\label{sec:exp:runtime}

On one core, ACL encodes $1$\,M joint-samples in $5.0$\,s and decodes them in $12.5$\,ms. The max-gate codec encodes in $24$\,s ($4.8\times$) and decodes in $0.48$\,s ($38\times$); the product encodes in $57$\,s ($12\times$) and decodes in $2.1$\,s ($168\times$, about $8{,}700$ frames per second of a $55$-joint skeleton), $0.76$\,s on four threads and $0.57$\,s on one GPU (pooled over all precisions). Decode time scales with the number of coded residuals; at $0.01$\,cm the median clip of a $93$-clip timing subset decodes in $0.10$\,s with the max-gate codec and $0.57$\,s with the product, the 95th percentile in $0.86$ and $4.5$\,s. The model weights take $870$\,KB (transformer) and $182$\,KB (MLP). Re-encoding the decoded clip into ACL (supplementary Sec.~E) costs one ACL compression, which dominates the load time, and a second lossy stage ($1.4$ to $1.8\times$ ACL's mean error), so a pipeline that must deliver ACL's error should run that stage at a tighter precision or keep the dense output; ACL at $p/2$ costs $10$ to $16\,\%$ more bytes, and we did not measure the composite error at that setting. Our equal-error comparisons are therefore claims about storage and transport, not about runtime.

\subsection{Failure Cases}
\label{sec:exp:failure}

\ifinlinefigs\input{sec/figs/fig_failure}\fi
\name{} gains least on short, low-rate clips of small skeletons (HumanAct12 $0.676\times$ at $0.01$\,cm, Edinburgh $0.628\times$, BFA $0.609\times$; Fig.~\ref{fig:exp:failure}), where the header and the model's cold start are amortized over few samples and few samples can be omitted. At loose precision, ACL's own misses on HiPHI widen both contracts, and the mean gate's guard, ten times ACL's clip mean, admits worst-case errors up to $24.7$\,cm at $1$\,cm on a handful of clips; we do not recommend the mean gate at $p \geq 0.3$\,cm. We do not compare against \prevname{}: its published numbers come from a different split with floating-point accounting and are listed in the supplement for context only.

%% file: sec/figs/fig_rd.tex
\begin{figure}[t]
    \centering
    \includegraphics[width=\columnwidth]{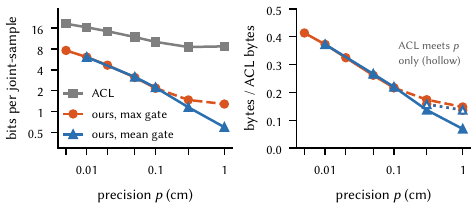}
    \caption{\capRD}
    \Description{\descRD}
    \label{fig:exp:rd}
\end{figure}

%% file: sec/figs/fig_storage.tex
\begin{figure}[t]
    \centering
    \includegraphics[width=\columnwidth]{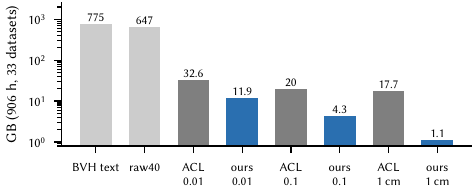}
    \caption{\capStorage}
    \Description{\descStorage}
    \label{fig:exp:storage}
\end{figure}

%% file: sec/figs/fig_qualitative.tex
\begin{figure*}[t]
    \centering
    \includegraphics[width=\textwidth]{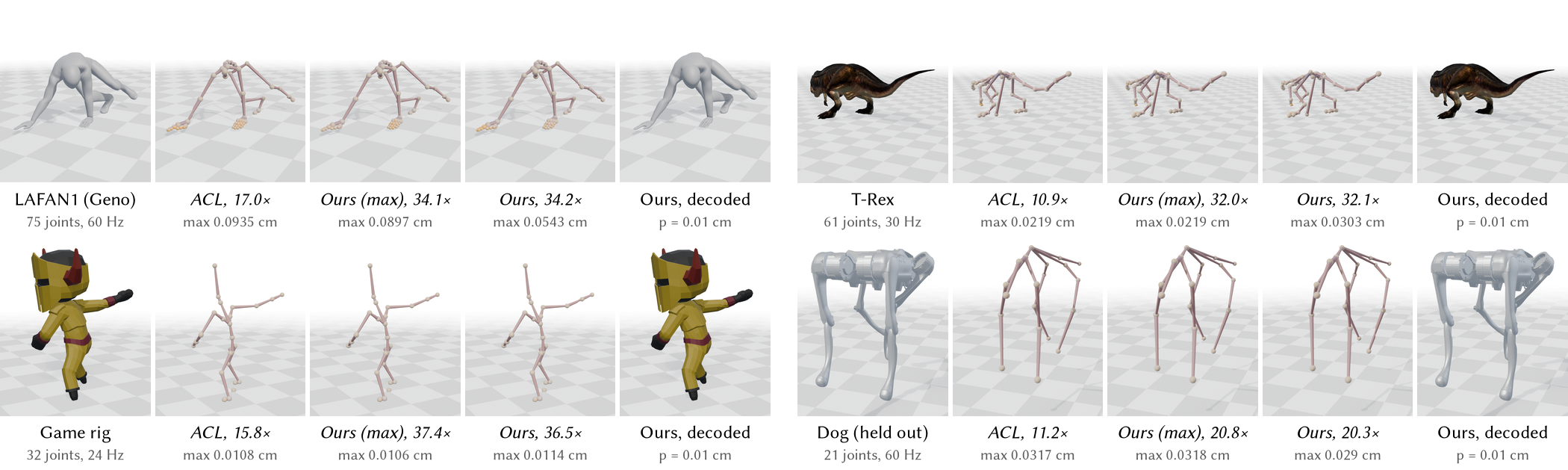}\\[2pt]
    \includegraphics[width=\textwidth]{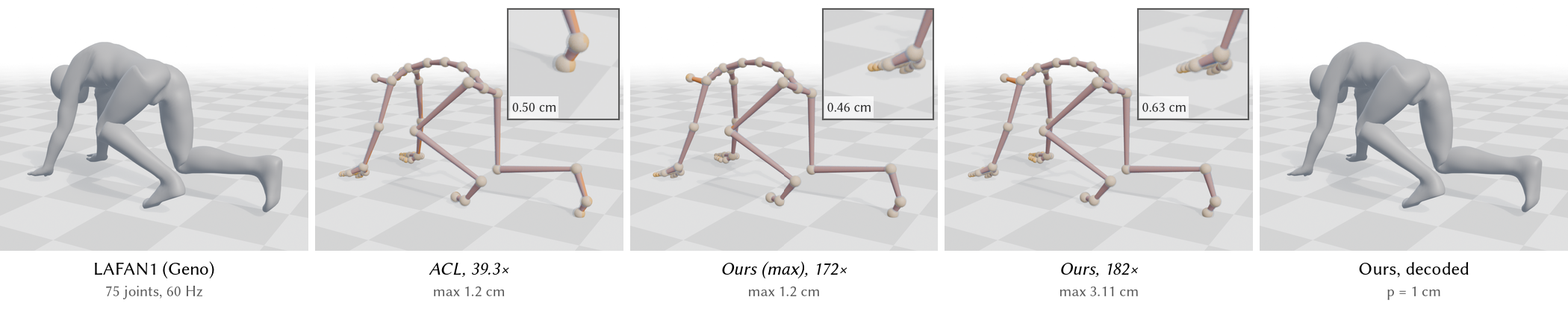}
    \caption{\capQualitative}
    \Description{\descQualitative}
    \label{fig:exp:qualitative}
\end{figure*}

%% file: sec/figs/fig_per_dataset.tex
\begin{figure*}[t]
    \centering
    \includegraphics[width=\textwidth]{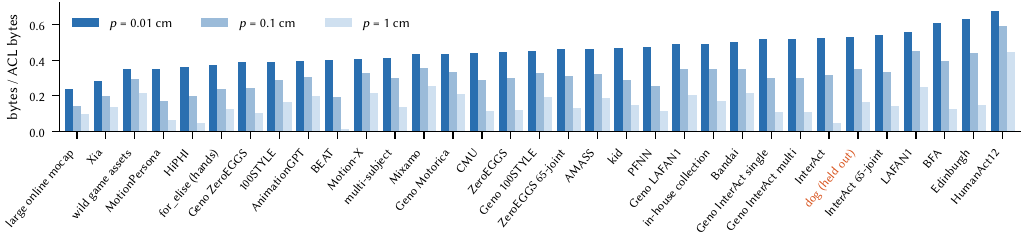}
    \caption{\capPerDataset}
    \Description{\descPerDataset}
    \label{fig:exp:dataset}
\end{figure*}

%% file: sec/figs/fig_failure.tex
\begin{figure}[t]
    \centering
    \includegraphics[width=\columnwidth]{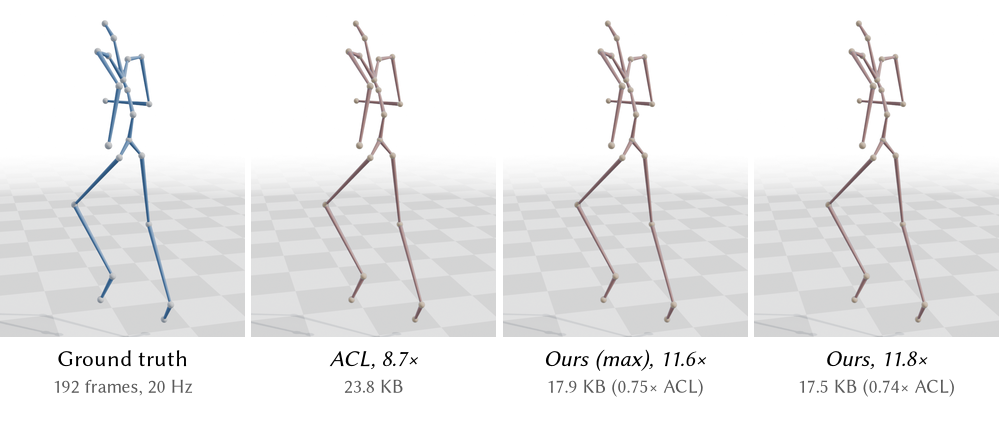}
    \caption{\capFailure}
    \Description{\descFailure}
    \label{fig:exp:failure}
\end{figure}

%% file: sec/7_conclusion.tex
\section{Limitations}
\label{sec:limitations}

\paragraph{No random access.}
\name{} decodes whole clips: the coder, predictors, and model step are sequential in time, so a single pose cannot be read in constant time, which ACL does in about half a microsecond, and a full clip takes $38$ to $168\times$ ACL's decompression time on one core, still far faster than real time (Sec.~\ref{sec:exp:runtime}). This is inherent to the design, not a matter of optimization: ACL is stateless and keeps the clip compressed in memory, decompressing only the poses a runtime's evaluation pass needs, whereas \name{} is a storage and distribution format decoded at load time, and re-encoding into ACL adds a second lossy stage. Encoding is $4.8$ to $12\times$ slower than ACL and runs offline, in parallel over clips.

\paragraph{Contracts relative to ACL}
Both contracts take their targets from ACL's result on the same clip, so the encoder needs ACL's statistics and inherits its misses at loose precision; the max gate holds ACL's worst case only within a $5\,\%$ tolerance, and errors are bounded at the sample times, not between them. A standalone contract in terms of $p$ costs $4$ to $6\,\%$ of the bytes on a test subset but was not run at corpus scale. At loose precision the max-gate encoder does not un-fold tracks (Sec.~\ref{sec:exp:contracts}); the fix is not part of the benchmarked codec.

\paragraph{Not validated as a representation.}
We motivated the codec by the hope that what it must send is closer to the motion's decisions than to the body, but we test it only as a codec. Neither the stream nor the entropy model is evaluated on a generative or recognition task, and the findings that interpolation, cross-joint structure, and transforms did not pay hold for the estimators, data, and precisions we tried, each with one architecture and training run; they do not prove that no such model can do better. Whether the codec can serve a learned motion model is future work.

\section{Conclusion}
\label{sec:conclusion}

We asked where the redundancy of skeletal motion lies and which part of a codec a learned model should take over, and answered by measurement. Predicting each quantized curve from its own past is the largest saving at tight precision; omitting samples, chosen per joint in closed loop through the hierarchy, is the second, and its share grows until the two are comparable at $1$\,cm. On the gaps such an encoder leaves there is little left to interpolate: none of the interpolators we tried, including a retrieval oracle over millions of training samples, beat linear interpolation, so the learned in-betweener at the core of \prevname{} had little left to learn. A network pays as the probability model of the residuals the codec must send. \name{} puts these findings into one skeleton-agnostic, bit-exact codec that needs $0.37\times$ ACL's bytes at ACL's default precision under ACL's own worst-case contract within tolerance, and $0.22\times$ at $0.1$\,cm at its mean error, with one model that serves every rig we tried, including a species absent from training. The two parts that carry the gain, each curve's slowly varying residual regime and the hierarchy-aware choice of which samples to send, are what a learned motion model would have to capture to work across bodies; building such a model on them is the next step.

\begin{acks}
We thank Nicholas Frechette, the author of ACL, for detailed discussions of ACL's
design goals and technical details. We thank Jun Xing, Tianshu Zhang and
Zhixin Piao for the very early discussions on motion compression. This work was partially funded by the Research Grants Council of Hong Kong (Ref: 17216826) and by the Innovation and Technology Commission of the HKSAR Government under the ITSP-Platform grant (Ref: ITS/335/23FP, ITS/469/24FP).
\end{acks}